\documentclass[aps, twocolumn, showpacs]{revtex4}
\begin{document}
\title{On the harmonic map equation}
\author { S. C. Tiwari\\
Institute of Natural Philosophy\\
c/o 1 Kusum Kutir Mahamanapuri,Varanasi 221005, India}
\begin{abstract}
A generalized harmonic map equation is presented based on the proposed action
functional in the Weyl space (PLA, 135, 315, 1989).
\end{abstract}
\maketitle

Let $M^p$ be a compact, orientable analytic manifold with coordinates $\{ x^\mu \}$, 
$\mu =1,2,... p$ and an analytic metric $g_{\mu\nu}$
\begin{equation}
ds^2 = g_{\mu\nu} dx^\mu dx^\nu
\end{equation}
and $N^q$ be another compact, analytic manifold with coordinates $\{ \phi ^a\}$,
$a=1,2,... q$ and an analytic metric $G_{ab}$
\begin{equation}
dl^2 = G_{ab} d\phi ^a d\phi ^b
\end{equation}
A mapping of class $C^2$ from $M^p$ to $N^q$
\begin{equation}
\phi :~ M \longrightarrow N,  x \longrightarrow \phi \{x\}
\end{equation}
represented in coordinates as $\phi ^a(x^\mu)$ is a harmonic mapping for the critical
points of an energy functional. The critical maps satisfy the Euler-Lagrange equations
\begin{equation}
\frac {1}{\sqrt {-g}} \frac {\partial {\sqrt {-g} g^{\mu \nu} \frac {\partial \phi^a}
{\partial x^\nu}}}{\partial x^ \mu} +\Gamma^{a}_{~bc} \frac {\partial \phi ^b}{\partial
 x^ \mu} \frac {\partial \phi ^c}{\partial x^ \nu} g^{\mu \nu} =0
\end{equation}
where $\Gamma ^{a}_{~bc}$ are the Christoffel symbols of the metric (2) on $N^q$.

Harmonic mapping provides a geometrical framework for field theories such as
Yang-Mills and sigma models \cite{1}. Singular behavior of the solutions of harmonic map
equation is of great interest in differential geometry and mathematical physics. In this
note I present a generalized harmonic map equation revisiting my work on Weyl 
geometry \cite{2} in the perspective of harmonic mapping.
Recently there has been a lot of activity in the higher curvature gravity \cite{3}, and
it would seem that my work on scalar field in gravity \cite{4} in which Brans-Dicke
theory, see \cite{5} for review, and nonlinear terms in the scalar curvature $R$ were
discussed would be relevant. There has been revival of Weyl geometry in both
cosmological \cite{6} and quantum mechanical \cite{7} contexts. In \cite{4} the
problem of matter energy conservation was also discussed. In the case of
unimodular relativity a modified covariant divergence law of stress tensor was
proposed \cite{8}. Recent work on the modification in matter energy conservation by Bock
\cite{9} seems significant; he also draws attention to similar but perhaps little known work 
of 1970s \cite{10}.

Let us consider the energy functional for harmonic mapping from $M^4$ to $N^1$
\begin{equation}
E ~=\frac {1}{2} \int \sqrt {-g}~g^{\mu \nu} \frac {\partial \phi}{\partial x^\mu}
\frac {\partial \phi}{\partial x^\nu}~d^4 x
\end{equation}
The new action functional in Weyl space postulating an in-invariant scalar field
$\psi$ proposed in \cite{2} reads
\begin{equation}
I ~= \int \sqrt {-g}~W~g^{\mu \nu} \frac {\partial \psi}{\partial x^\mu}
\frac {\partial \psi}{\partial x^\nu}~d^4 x
\end{equation}
Here $W$ is scalar curvature in Weyl space. The field equation for $\psi$ obtained from
this action (6) is 
\begin{equation}
\frac {W}{\sqrt {-g}} \frac {\partial {\sqrt {-g} g^{\mu \nu} \frac {\partial \psi}
{\partial x^\nu}}}{\partial x^ \mu}+\frac {\partial W}{\partial
 x^ \mu} \frac {\partial \psi}{\partial x^ \nu} g^{\mu \nu} =0
\end{equation}
On the other hand the harmonic map equation from the energy functional $E$ (5) is
\begin{equation}
\frac {1}{\sqrt {-g}} \frac {\partial {\sqrt {-g} g^{\mu \nu} \frac {\partial \phi}
{\partial x^\nu}}}{\partial x^ \mu}=0
\end{equation}
Comparing expressions (7) and (8) it is evident that for constant $W$ both are
equivalent. Of course, $W$ being a scalar curvature with Weyl power 2 or a co-scalar
geometrically one has to be careful in setting this condition i.e. $W=constant$ in
Weyl space. However Eq.(7) can be viewed as a generalized harmonic map equation.
In general, the generalized harmonic map equation from $M^p$ to $N^q$ is obtained to be
\begin{equation}
\frac {R}{\sqrt {-g}} \frac {\partial {\sqrt {-g} g^{\mu \nu} \frac {\partial \phi^a}
{\partial x^\nu}}}{\partial x^ \mu} +R~\Gamma^{a}_{~bc} \frac {\partial \phi ^b}{\partial
 x^ \mu} \frac {\partial \phi ^c}{\partial x^ \nu} g^{\mu \nu} +g^{\mu \nu} \frac 
{\partial {\phi ^a}}{\partial {x^\mu}}~\frac {\partial R}{{\partial {x^\nu}}}=0
\end{equation}
Note that $R$ is the scalar curvature of the manifold $M^p$, and there is no
Weyl structure in this manifold.

I tried to investigate Yamabe's problem and conformal mapping using the generalized
harmonic map equation, but there is no definitive result so far. It may be of interest
to explore it for the higher curvature theories.

I thank Prof. S. D. Odintsov for \cite{3} and Dr. Triloki Singh for \cite{5}.
The Library facility at Banaras Hindu University is acknowledged.

\end{document}